\documentclass[bimj,fleqn]{w-art}
\usepackage{times}
\usepackage{w-thm}
\usepackage[authoryear]{natbib}
 \usepackage{amsmath, amssymb}
 \usepackage{graphicx, subfigure} 
 \usepackage{enumerate}
 \usepackage{url, hyperref} 
 \usepackage{setspace}
\usepackage{xcolor}
\usepackage{float}

\newcommand{\revisioncolor}{black}
\newcommand{\mytextcolor}[1]{\textcolor{\revisioncolor}{#1}}

\chardef\bslash=`\\ 

\newcommand{\bY} {\mathbf {Y}}

\newcommand{\bs} {\mathbf {s}}

\newcommand{\logit} {\textrm{logit}}
\newcommand{\Binomial} {\textrm{Binomial}}
\newcommand{\Beta} {\textrm{Beta}}

\numberwithin{equation}{section}
\numberwithin{table}{section}

\begin{document}

\DOIsuffix{bimj.200100000}
\Volume{xx}
\Issue{xx}
\Year{2025}
\pagespan{1}{}
\keywords{Bayesian basket design; local power prior, dynamic borrowing, oncology trials\\[1pc]
\noindent\hspace*{-4.2pc} 
Supporting Information for this article is available from the author or on the WWW under\break \hspace*{-4pc} \underline{http://dx.doi.org/10.1022/bimj.XXXXXXX}\\[1pc]
\noindent\hspace*{-4.2pc} 
\textit{This reprint is in press at Biometrical Journal and may differ from the published version in typographic detail.}
}  

\title[Bayesian Basket Trial]{A Bayesian Basket Trial Design Using Local Power Prior}

\author[Zhou {\it{et al.}}]{Haiming Zhou\footnote{Corresponding author; {\sf{E-mail: haiming.zhou@daiichisankyo.com}}}\inst{,1}} 
\address[\inst{1}]{Daiichi Sankyo, Inc, USA}
\author[]{Rex Shen\inst{2}}
\address[\inst{2}]{Stanford University, USA}
\author[]{Sutan Wu\inst{1}}
\author[]{Philip He\inst{1}}

\Receiveddate{zzz} \Reviseddate{zzz} \Accepteddate{zzz} 

\begin{abstract}
In recent years, basket trials, which allow the evaluation of an experimental therapy across multiple tumor types within a single protocol, have gained prominence in early-phase oncology development. Unlike traditional trials, which evaluate each tumor type separately and often face challenges with limited sample sizes, basket trials offer the advantage of borrowing information across various tumor types to enhance statistical power. However, a key challenge in designing basket trials is determining the appropriate extent of information borrowing while maintaining an acceptable type I error rate control. In this paper, we propose a novel 3-component local power prior (local-PP) framework that introduces a dynamic and flexible approach to information borrowing. The framework consists of three components: global borrowing control, pairwise similarity assessments, and a borrowing threshold, allowing for tailored and interpretable borrowing across heterogeneous tumor types. Unlike many existing Bayesian methods that rely on computationally intensive Markov chain Monte Carlo (MCMC) sampling, the proposed approach provides a closed-form solution, significantly reducing computation time in large-scale simulations for evaluating operating characteristics. Extensive simulations demonstrate that the proposed local-PP framework performs comparably to more complex methods while significantly shortening computation time. 
\end{abstract}
 
\maketitle

\section{Introduction}
In recent years, basket trials have become a key innovation in early-phase oncology, allowing for the evaluation of targeted therapies across multiple tumor types, referred to as “baskets,” that share a common molecular alteration or biomarker within a single protocol. This represents a fundamental shift from traditional approaches that focus on the tumor’s tissue of origin, moving towards molecular characteristics-based drug development. Since the FDA’s accelerated approval of pembrolizumab in unresectable/metastatic MSI-H or dMMR solid tumors in 2017, six additional indications have been approved by the FDA (Appendix Table~A1). 

Most basket trials are conducted in exploratory settings, where the primary objective is to identify the cancer types for which an experimental drug shows promising activity for subsequent phases of development. For instance, the clinical study of BRAF V600 mutated tumors \citep{hyman2015vemurafenib} was conducted as an exploratory basket trial to assess the preliminary efficacy of vemurafenib across six pre-specified cancer types: non-small cell lung cancer (NSCLC), colorectal cancer treated with vemurafenib (CRC vemu), CRC treated with vemurafenib and cetuximab (CRC vemu+cetu), cholangiocarcinoma (bile duct), Erdheim-Chester disease or Langerhans' cell histiocytosis (ECD or LCH), anaplastic thyroid cancer (ATC), and colorectal cancer (CRC). In contrast to traditional trials, which typically investigate each cancer types separately in phase 2 studies, basket trials provide opportunities to borrow information across baskets, thereby improving trial efficiency, particularly when some tumor types have small sample sizes. However, determining the optimal level of borrowing across potentially heterogeneous tumor types remains a challenge.

Multiple Bayesian approaches have been developed to address this challenge. One of the earliest was the Bayesian hierarchical model (BHM) by \cite{berry2013bayesian}, which assumes that patients across different baskets respond to the therapy homogeneously. Since then, more flexible methods have been proposed to account for potential heterogeneity in treatment effects across baskets, such as exchangeability–nonexchangeability (EXNEX) \citep{neuenschwander2016robust}, Bayesian cluster hierarchical model (BCHM) \citep{chen2020bayesian}, multisource exchangeability modeling (MEM) \citep{hobbs2018bayesian}, Robust Bayesian Hypothesis Testing (RoBoT) \citep{zhou2021robot}, and multiple cohort expansion (MUCE) \citep{lyu2023muce}; see \citep{pohl2021categories} for a comprehensive review of Bayesian basket trial design methods. Most of these methods depend on Markov Chain Monte Carlo (MCMC) sampling for posterior inference, therefore can be computationally intensive, especially when exploring the operating characteristics in large-scale simulations. To address the computational challenges, several MCMC-free methods have been proposed. These include Simon’s two-stage basket trial design \citep{simon2016bayesian}, information borrowing based on Jensen-Shannon divergence \citep{fujikawa2020bayesian}, Bayesian model averaging (BMA) \citep{psioda2021bayesian}, and local-MEM \citep{liu2022bayesian}. While these methods improve computational efficiency, they have limitations in terms of flexibility and interpretability. For example, Simon’s design either borrows fully among all baskets or does not borrow at all, while BMA and local-MEM require extensive computation as the number of baskets increases. Fujikawa’s method offers a computational advantage, but its tuning parameters are not easily interpretable. Recently, \cite{baumann2024basket} explored the use of power prior designs in basket trials, with a focus on weight specification as a key factor in controlling information borrowing. The power prior approach \cite{ibrahim2015power} incorporates information from other baskets through a weighted likelihood. In the setting of borrowing historical data, the power prior approach constructs an informative prior by incorporating historical data through a weight parameter (often called the power parameter), which is typically pre-specified (e.g., set to 0.5) based on subjective judgement. This weight adjusts the extent of borrowing, with greater weight assigned when historical data are more relevant. The power prior concept can naturally be extended to the basket trials, where the weight between baskets is determined according to their similarity in the endpoint such as objective response. The advantage of the power prior framework lies in its significantly shorter computation time due to its closed form posterior distribution and its clear interpretation of the borrowing mechanisms. \cite{gravestock2019power} applied empirical Bayes (EB) methods for weight estimation in multiple historical study settings, and \cite{baumann2024basket} extended this approach to basket trial designs.

In this paper, we propose a novel 3-component local power prior (local-PP) framework that advances the flexibility and interpretability of the power prior approach. Our framework consists of three components: global borrowing control ($a$), pairwise similarity assessments ($s_{ij}$), and a borrowing threshold ($\Delta$). The global control parameter $a$ governs the overall extent of borrowing across all baskets, while the pairwise similarity parameter $s_{ij}$ assesses the similarity between specific tumor types. The threshold parameter $\Delta$ limits borrowing when substantial differences in response rates are observed between baskets. Unlike previous methods that rely heavily on MCMC, our local-PP framework allows for a closed-form solution to the posterior distribution, making it computationally efficient for large-scale simulations. Additionally, the framework accommodates unequal samples sizes across baskets, ensuring that borrowing is proportional to the available information in each basket, thus preserving the integrity of the overall analysis. Moreover, the 3-component framework allows for dynamic control at both the global and local levels, offering a more tailored approach to borrowing. Although \cite{baumann2024basket} explored similar concepts in power prior designs, our approach goes further by integrating these components into a unified framework that is easily interpretable and adaptable based on specific trial needs.

The rest of this paper is organized as follows: Section~\ref{sec:methods} introduces the local-PP framework and its three components. Section~\ref{sec:simulation} presents a comprehensive simulation study comparing our method with existing approaches. In Section~\ref{sec:example}, we apply our method to a case study for BRAF V600 mutated rare cancers, demonstrating its practical utility in a real-world context. Finally, Section~\ref{sec:discussion} concludes with a discussion of the practical implications and potential extensions of our framework.

\section{Methods}\label{sec:methods}
Consider a basket trial with $B$ tumor types (i.e., baskets). Let $p_i$ denote the ORR for basket $i$. Suppose $n_i$ patients enrolled in basket $i$ and $Y_i$ of them achieved tumor response. Then
$$
Y_i | p_i \sim \text{Binomial}(n_i, p_i), \quad i = 1, \ldots, B.
$$
The mechanism of information borrowing across baskets is facilitated by incorporating informative priors on $p_i$ or $\logit(p_i)$, e.g., BHM \citep{berry2013bayesian}, EXNEX \citep{neuenschwander2016robust}, and MEM \citep{hobbs2018bayesian}. \mytextcolor{Unlike} many methods that require Markov chain Monte Carlo (MCMC) sampling for posterior inference, we introduce a local power prior method that eliminates \mytextcolor{this requirement}. This is particularly advantageous for practical use in exploring trial design operating characteristics by simulations.

\subsection{Power Prior}\label{sec:pp}
The power prior approach models basket $i$ data while using an informative prior constructed from the other baskets. Let $\pi(\cdot)$ be a generic notation for the density function of a random variable. The power prior \citep{ibrahim2015power} for $p_i$ is constructed below:
\begin{equation} \label{eq:PP}
\pi(p_i | Y_j \text{ for } j \neq i) \propto \pi(p_i | b_{1i}, b_{2i}) \times \prod_{j \neq i} \pi(Y_j | n_j, p_i)^{w_{ij}},
\end{equation}
where $\pi(\cdot | b_1, b_2)$ is the density function of $\Beta(b_1, b_2)$, $\pi(\cdot | n, p)$ is the probability mass function of $\Binomial(n, p)$, $w_{ij}$ is the power parameter interpreted as the amount of borrowing from basket $j$, and $(b_{1i}, b_{2i})$ are pre-specified hyperparameters of the initial beta prior for $p_i$. When $w_{ij} = 0$, this prior reduces to the hyperprior without borrowing any information from other baskets. The power prior approach achieves significant computational advantage over BHM-based methods because the posterior distribution of $p_i$ has a closed form of beta distribution:
\begin{equation}\label{eq:posterior}
p_i | \bY, b_{1i}, b_{2i} \sim \text{Beta}\left(b_{1i} + Y_i + \sum_{j \neq i} w_{ij} Y_j, b_{2i} + n_i - Y_i + \sum_{j \neq i} w_{ij} (n_j - Y_j)\right),
\end{equation}
where $\bY = (Y_1, \ldots, Y_B)$. Let $\Omega$ denote a $B \times B$ matrix with $ij$-th element being $w_{ij}$ and all diagonal elements being ones. The weight parameter $w_{ij}$ has an explicit interpretation. For example, $w_{ij} = 0.4$ indicates that we borrow 40\% of information from basket $j$ when evaluating basket $i$. {Several relevant methods have been proposed to determine $w_{ij}$, including MEM \citep{hobbs2018bayesian}, Jensen-Shannon divergence \citep{fujikawa2020bayesian} and local-MEM \citep{liu2022bayesian}. We provide a brief review of these methods below.} 

\textbf{MEM.} The MEM method assumes that $w_{ij} = w_{ji} \in \{0, 1\}$ with value 1 (0) indicating that baskets $i$ and $j$ are exchangeable (independent), leading to $J = \prod_{i=1}^{B-1} 2^i$ possible model configurations. Each $w_{ij}$ is assumed to follow a Bernoulli prior with $P(w_{ij} = 1) = 0.5$. The posterior distribution of $p_i$ is derived by averaging over the posterior distribution of $\{w_{i1}, \ldots, w_{iB}\}$. The R package \texttt{basket} \citep{kane2020analyzing} provides two methods to conduct the posterior inference: the exact method which enumerates all model configurations, and the MCMC sampling method formulated from the Metropolis algorithm. The exact method is only computationally feasible for $B < 7$ and the MCMC method can be time-consuming in the large-scale simulations due to extensive posterior samplings.

\textbf{Jensen-Shannon Divergence (JSD).} Denote $f_i(\cdot)$ as the posterior density function for $p_i$ based on basket $i$ data only, i.e., Beta $(b_{1i} + Y_i, b_{2i} + n_i - Y_i)$. \cite{fujikawa2020bayesian} proposed an approach based on Jensen-Shannon divergence \citep{fuglede2004jensen}. For baskets $i$ and $j$, denote
\begin{equation}\label{eq:JSD}
    w_{ij}^* = 1 - \text{JS}(f_i, f_j) = 1 - \frac{1}{2} \left( \text{KL}\left(f_i \| \frac{f_i + f_j}{2}\right) + \text{KL}\left(f_j \| \frac{f_i + f_j}{2}\right) \right),
\end{equation}
where $\text{KL}(f_i \| f_i') = \int_0^1 f_i(x) \log \frac{f_i(x)}{f_i'(x)} dx$ is the Kullback-Leibler divergence between densities $f_i(\cdot)$ and $f_i'(\cdot)$. The resulted $w_{ij}^*$ ranges from 0.307 to 1. To allow for weaker or no borrowing among dissimilar baskets, \cite{fujikawa2020bayesian} proposed to set $w_{ij} = {w_{ij}^{*}}^\epsilon \mathbb{I}({w_{ij}^{*}}^\epsilon > \tau)$, where $\epsilon \geq 1$ is a power tuning parameter and $\tau \in [0, 1]$ is a threshold tuning parameter. They recommend setting $\epsilon = 2$ and trying different $\tau$ values from [0, 0.5].

\textbf{Local-MEM.} This method considers all possible partitions of the $B$ baskets into clusters of varying sizes with each partition corresponding to each configuration of $\Omega$. For example, one can form $L = 5$ possible partitions of $B = 3$ baskets into clusters \mytextcolor{$\{1, 2, 3\}$, $\{(1, 2), 3\}$, $\{(1, 3), 2\}$, $\{(2, 3), 1\}$} and $\{(1, 2, 3)\}$, where under each partition, set $w_{ij} = 1$ if baskets $i$ and $j$ are in the same cluster and 0 otherwise. Denote $\{\Omega_1, \ldots, \Omega_L\}$ as the collection of all possible configurations of $\Omega$. Comparing to the original MEM method, the local-MEM method does not allow information borrowing across different clusters. To determine which $\Omega$ configuration to use via posterior inference, \cite{liu2022bayesian} assumed the following prior
$$
\pi(\Omega_j) = \frac{|\Omega_j|^\delta}{\sum_{j=1}^L |\Omega_j|^\delta}, \quad j = 1, \ldots, L,
$$
where $|\Omega_j|$ denote the number of clusters under the configuration $\Omega_j$, and $\delta$ is a tuning parameter with larger positive $\delta$ values favoring partitions with more clusters. \cite{liu2022bayesian} investigated the prior effect by considering $\delta = 0, 1, 2$. Let $\Omega^*$ denote the partition with the largest posterior probability and its posterior probability is denoted as $\pi(\Omega^*)$. The local-MEM method sets $w_{ij} = \pi(\Omega^*)$ if baskets $i$ and $j$ are in the same cluster under configuration $\Omega^*$ and 0 otherwise. When the number of baskets is large (say $B > 7$), this method can become computationally intensive.

\subsection{Dynamic Borrowing Mechanism}\label{sec:borrowing}
From equation~\ref{eq:posterior}, the posterior effective sample size (ESS) for basket $i$, as described in \cite{hobbs2018bayesian}, is $ESS_{1i} = n_i + b_{1i} + b_{2i} + \sum_{k \neq i} w_{ik} n_k$ with borrowing, and is $ESS_{0i} = n_i + b_{1i} + b_{2i}$ without borrowing (i.e. when $w_{ij}=0$ for all $k\neq i$). To quantify the extent of borrowing at the basket level, we define the borrowing factor (BF) for basket $i$ as 
\begin{equation}\label{eq:bf}
BF_i = \frac{ESS_{1i} - ESS_{0i}}{n_i} = \frac{\sum_{k \neq i} w_{ik} n_k}{n_i}.
\end{equation}
Here, $BF_i$ can be interpreted as the equivalent number of subjects borrowed from other baskets, relative to the sample size of basket $i$. For instance, if $BF_i=2$, the equivalent number of subjects borrowed from other baskets is twice the sample size of basket $i$. Generally, higher BF values are associated with a greater risk of type I error inflation. To better control the maximum allowable borrowing in terms of BF, we propose decomposing the weight parameter $w_{ij}$ into three components:
\begin{equation}\label{eq:weight}
w_{ij} = \min\left(a\frac{ n_i}{n_{-i}}, 1\right) \cdot s_{ij} \cdot \mathbb{I}\left(\left|\hat{p}_i - \hat{p}_j\right| < \Delta\right),
\end{equation}
{where $\hat{p}_i=\frac{Y_i}{n_i}$, $\hat{p}_j=\frac{Y_j}{n_j}$, and $n_{-i} = \sum_{k\neq i} n_k$ presents the total sample size for all baskets except basket $i$, $a\geq 0$ is a discounting parameter that controls the overall amount of borrowing in terms of BF across all baskets}, $s_{ij}$ is a similarity parameter quantifying the degree of borrowing from basket $j$ to basket $i$, and $\Delta \in [0,1]$ is a threshold parameter allowing borrowing from basket $j$ only when the observed difference in ORR between baskets $i$ and $j$ is below the threshold. To encourage borrowing, we recommend avoiding $\Delta$ values smaller than 0.1 and suggest selecting $\Delta$ based on the null and alternative hypotheses, clinical considerations, and \mytextcolor{simulation-based evaluations}. 

According to equation~\ref{eq:bf}, a smaller $a$ results in less borrowing in terms of the borrowing factor a priori. When $a=0$, the model reduces to the independent model without borrowing. Conversely, when $a=\max\{ n_{-i}/n_i: i=1,\ldots,B \}$, there is no global discount for borrowing across tumor baskets, aside from the effects of $s_{ij}$ and $\Delta$, which may lead to considerable type I error inflation. Using the weight parameter $w_{ij}$ defined in equation~\ref{eq:weight}, we have $BF_i \leq \min\{a, n_{-i}/n_i\}\leq a$, meaning that $a$ can be interpreted as the maximum allowable equivalent number of subjects borrowed from other baskets. For instance, with 5 tumor baskets and each having 40 subjects, setting $a=0.5$ limits the maximum borrowing for tumor basket 1 from the other 4 to $40\times 0.5 = 20$ equivalent subjects. This interpretation provides guidance for selecting an appropriate range of $a$ values. A particular choice of $a=1$ implies that the maximum allowable number of subjects borrowed from other baskets is equal to the current basket's sample size, which can serve as a reasonable starting point for optimizing $a$. Further refinements should be explored through simulations, as discussed in Section~\ref{sec:simulation}. 

Regarding the determination of $s_{ij}$, we propose estimating them using an empirical Bayes (EB) approach by maximizing their marginal likelihoods. To isolate the effect of $s_{ij}$, we exclude the other two weight components, $a$ and $\Delta$, during the empirical Bayes estimation. The $s_{ij}$ values are estimated based solely on data from baskets $i$ and $j$, using the following model:
$$
Y_i | p_i \sim \text{Binomial}(n_i, p_i)
$$
$$
\pi(p_i | s_{ij}) \propto \pi(p_i | b_{1i}, b_{2i}) \pi(Y_j | n_j, p_i)^{s_{ij}}.
$$
Then, the marginal likelihood of observing $Y_i$ given $Y_j$ and $s_{ij}$ is
$$
L(Y_i | Y_j, s_{ij}) = \frac{\int_0^1 \pi(Y_i | n_i, p_i) \pi(p_i | b_{1i}, b_{2i}) \pi(Y_j | n_j, p_i)^{s_{ij}} dp_i}{\int_0^1 \pi(p_i | b_{1i}, b_{2i}) \pi(Y_j | n_j, p_i)^{s_{ij}} dp_i}.
$$
It can be shown that $L(Y_i | Y_j, s_{ij})$ is proportional to $m(s_{ij})$ given by
\begin{equation}\label{eq:EBpairwise}
    m(s_{ij}) = \frac{\text{Be}(b_{1i} + Y_i + s_{ij} Y_j, b_{2i} + n_i - Y_i + s_{ij} (n_j - Y_j))}{\text{Be}(b_{1i} + s_{ij} Y_j, b_{2i} + s_{ij} (n_j - Y_j))},
\end{equation}
where $\text{Be}(b_1,b_2) = \int_0^1 t^{b_1-1} (1-t)^{b_2-1} dt$ is the beta function with parameters $b_1$ and $b_2$. The parameters $s_{ij}$ can be estimated by maximizing $m(s_{ij})$ independently for all $i \neq j$. For example, suppose the observed data are $(Y_1, \ldots, Y_5) = (2,9,11,13,20)$ and $(n_1, \ldots, n_5) = (25, 25, 25, 25, 25)$, and set the beta hyperprior with $b_{1i} = b_{2i} = 0.5$. Then the estimated weights $s_{ij}$ are:
\begin{center}
\begin{tabular}{c|ccccc}
Basket & 1 & 2 & 3 & 4 & 5 \\
\hline
  1 & 1.00 & 0.04 & 0.02 & 0.00 & 0.00 \\ 
  2 & 0.06 & 1.00 & 1.00 & 0.58 & 0.02 \\ 
  3 & 0.04 & 1.00 & 1.00 & 1.00 & 0.05 \\ 
  4 & 0.02 & 0.57 & 1.00 & 1.00 & 0.10 \\ 
  5 & 0.00 & 0.02 & 0.04 & 0.09 & 1.00 \\ 
\end{tabular}
\end{center}

Alternatively, we can treat basket $i$ as the current data and all other baskets as multiple historical datasets, and estimate $\bs_{-i}= \{s_{ij}, j\neq i\}$ globally by maximize its marginal likelihood, given by
\begin{equation}\label{eq:EBglobal}
    m(\bs_{-i}) = \frac{\text{Be}(b_{1i} + Y_i + \sum_{j\neq i} s_{ij} Y_j, b_{2i} + n_i - Y_i + \sum_{j\neq i} s_{ij} (n_j - Y_j))}{\text{Be}(b_{1i} + \sum_{j\neq i}s_{ij} Y_j, b_{2i} + \sum_{j\neq i}s_{ij} (n_j - Y_j))}.
\end{equation}
We refer to the method based on maximizing equation~\ref{eq:EBpairwise} as pairwise empirical Bayes (PEB) and the method based on maximizing equation~\ref{eq:EBglobal} as global empirical Bayes (GEB). Using the same example as before, the resulting GEB weights $s_{ij}$ are:
\begin{center}
\begin{tabular}{c|ccccc}
Basket & 1 & 2 & 3 & 4 & 5 \\
\hline
  1 & 1.00 & 0.04 & 0.00 & 0.00 & 0.00 \\ 
  2 & 1.00 & 1.00 & 1.00 & 1.00 & 0.12 \\ 
  3 & 1.00 & 1.00 & 1.00 & 1.00 & 1.00 \\ 
  4 & 0.12 & 1.00 & 1.00 & 1.00 & 1.00 \\ 
  5 & 0.00 & 0.00 & 0.00 & 0.09 & 1.00 \\ 
\end{tabular}
\end{center}

We observe that the GEB weights lead to non-intuitive borrowing behaviors. For instance, basket 3 borrows 100\% from all other baskets, even though baskets 1 and 5 have significant different ORRs compared to basket 3. This occurs because, in GEB, baskets 1 and 5 are treated as part of a pooled historical data, resulting in a combined ORR that matches basket 3. For this reason, we generally do not recommend using unadjusted GEB weights in basket trials. However, our proposed 3-component framework mitigates this non-intuitive borrowing issue, as demonstrated in the similarity matrix betlow, where the weights $w_{ij}$ are adjusted by equation~\ref{eq:weight} with $a=1$ and $\Delta=0.3$.
\begin{center}
\begin{tabular}{c|ccccc}
Basket & 1 & 2 & 3 & 4 & 5 \\
\hline
  1 & 1.00 & 0.01 & 0.00 & 0.00 & 0.00 \\ 
  2 & 0.25 & 1.00 & 0.25 & 0.25 & 0.00 \\ 
  3 & 0.00 & 0.25 & 1.00 & 0.25 & 0.00 \\ 
  4 & 0.00 & 0.25 & 0.25 & 1.00 & 0.25 \\ 
  5 & 0.00 & 0.00 & 0.00 & 0.02 & 1.00 \\ 
\end{tabular}
\end{center}

\cite{gravestock2019power} compared the unadjusted PEB and GEB weights in the multiple historical study setting for binary outcomes, demonstrating that GEB exhibited superior operating characteristics in their simulations. Recently, \cite{baumann2024basket} applied the unadjusted GEB weights in the context of basket trials, using the same power prior as in equation~\ref{eq:PP}, and compared it with several other methods (excluding PEB) for deriving $w_{ij}$, recommending GEB weights when controlling type I error inflation is a key concern. In this paper, we compare PEB with GEB after adjusting the weights using the proposed 3-component framework, focusing on their operating characteristics in terms of type I error control and power. We refer to the method that uses the power prior in equation~\ref{eq:PP} along with the proposed 3-component framework in equation~\ref{eq:weight} as local-PP, and we denote the local-PP method using PEB (GEB) weights as local-PP-PEB (local-PP-GEB).

\subsection{Type I Error and Calibration}\label{sec:Q}
Suppose we would like to enroll up to $n_i$ patients for tumor type $i$ and conduct $K$ interim futility analyses when the sample size reaches $n_{i1} < n_{i2} < \ldots < n_{iK} < n_i$ and one final analysis when the sample size reaches the maximum $n_i$. Let $Y_{ik}$ denote the number of responses at the $k$-th interim analysis for basket $i$, we stop the accrual to basket $i$ and claim futility if $Y_{ik} \leq r_{ik}$, where $r_{ik}$ is a pre-specified futility boundary. When all baskets have either enrolled the maximum number of patients or stopped enrollment due to futility, we perform the final analysis. Let $A$ denote the set of baskets included in the final analysis that were not deemed futile at interim, and let $D_i$ present the accumulated data for basket $i$ at the final analysis. Tumor type $i \in A$ is claimed promising if $P(p_i > p_{0} | D_i, i \in A) > Q_i$, where $p_0$ is a pre-specified non-promising ORR and $Q_i$ is a pre-specified efficacy cutoff.

The efficacy cutoffs $Q_i$ can be calibrated via simulations to control the type I error rate for each basket at a desired level given a specific sample size $n_i$. A smaller efficacy cutoff increases power but also inflates type I error rate. Conversely, the sample size can be determined based on the pre-specified $Q_i$ and power. In early-phase oncology trials, futility interim is a common practice to enable early termination of ineffective experimental treatment. The futility stopping boundaries $r_{ik}$ can be determined using various statistical approaches, such as the Bayesian optimal phase 2 (BOP2) design \citep{zhou2017bop2}. Additionally, the interim futility stopping boundaries are integrated into the calibration of $Q_i$.

In basket trials with multiple tumor types, various types of type I error can be considered including the basket-wise type I error rate (BWER) \citep{liu2022bayesian}, family-wise type I error rate (FWER) \citep{zhou2021robot}, false positive rate (FPR) which is the average of BWERs \citep{jiang2021optimal}, and the false discovery rate (FDR) \citep{zabor2022bayesian} which presents the portion of false positives among the claimed promising baskets. Since our focus here is on exploratory early-phase studies, we recommend calibrating $Q_i$ is based on BWER at a desired level $\alpha$ (e.g., 0.1) under the global null scenario (i.e., $p_i = p_{0}$ for all $i$), without consideration of multiplicity adjustment. The detailed calibration method by simulation is described in Appendix A2.

\subsection{Tuning and Performance Evaluation}\label{sec:tuning}
Section~\ref{sec:borrowing} provides general considerations for selecting the global borrowing parameter $a$ and the threshold borrowing parameter $\Delta$ within the context of the proposed local-PP method. As with other basket design models, it is unlikely to provide the universal choices for both parameters. We advocate for an optimization approach based on enumerated trial scenarios of interest, ranging from a global null scenario (where no baskets show promise) to a global alternative scenario (where all baskets show promise). To evaluate performance across these specified scenarios, metrics such as average basket-wise type I error, basket-wise power, true positive rate (TPR) which is an average of basket-wise power, and correct classification rate (CCR) can be used. \cite{broglio2022comparison} utilized the average CCR across specified scenarios to compare several BHM-based methods. In the following section, we adopt these evaluation measures to evaluate the performance of the proposed local-PP method in comparison to other relevant methods.

\section{Simulation study} \label{sec:simulation}
\subsection{Scenario Settings} \label{sec:scenarios}
Consider a design with $B=5$ tumor types. Suppose the non-promising ORR under the null hypothesis is $p_{0} = 0.15$ and the target ORR is $p_{1} = 0.30$ for all tumor types. The maximum sample size for each basket is $n_i = 25$, with one interim futility analysis conducted after the first 10 subjects: stop basket $i$ if the number of responses is less than or equal to 1. The stopping boundary of 1 is determined using the BOP2 design, which yields an approximate 15\% early stopping rate when the ORR is $0.30$ and about 54\% when the ORR is $0.15$. A total of six scenarios are considered for comparing various methods, as described in Table~\ref{tab:scenarios}. Scenario S1 represents the global null, with $Q_i$ calibrated in this scenario to ensure the BWER is controlled at $\alpha=0.1$. Scenario S6 presents a global alternative. Scenarios S2-S5 involve heterogeneous ORRs across the baskets. For each scenario, we simulate $M=5,000$ trials. Each simulated trial first \mytextcolor{undergoes} an interim futility assessment for each basket, and only those baskets with more than \mytextcolor{one response} proceed to the final analysis.  \mytextcolor{With $M=5,000$ replicates, differences at the third decimal place in reported proportions should be interpreted with caution, as they may fall within Monte Carlo variability.}

\begin{table}[h]
\centering
\caption{Simulation scenarios}
\label{tab:scenarios}
\begin{tabular}{cccccc}
\hline
Scenario & Basket 1 & Basket 2 & Basket 3 & Basket 4 & Basket 5 \\
\hline
S1 & 0.15 & 0.15 & 0.15 & 0.15 & 0.15 \\
S2 & 0.15 & 0.15 & 0.15 & 0.30 & 0.30 \\
S3 & 0.15 & 0.30 & 0.30 & 0.30 & 0.30 \\
S4 & 0.15 & 0.30 & 0.30 & 0.45 & 0.45 \\
S5 & 0.15 & 0.45 & 0.45 & 0.45 & 0.45 \\
S6 & 0.30 & 0.30 & 0.30 & 0.30 & 0.30 \\
\hline
\end{tabular}
\end{table}

\subsection{Models Specifications} \label{sec:sim:spec}
\mytextcolor{To ensure a fair evaluation of the proposed local-PP methods, we compare them with several established approaches for basket trial designs:}
\begin{itemize}\itemsep0em
    \item \mytextcolor{Independent model (IM): The power prior method with $\{w_{ij}=0: i\neq j\}$, meaning no borrowing occurs across baskets.}
    \item \mytextcolor{PP-PEB: The power prior method with unadjusted PEB weights, as defined in Section~\ref{sec:borrowing}.}
    \item \mytextcolor{PP-GEB: The power prior method with unadjusted GEB weights, as defined in Section~\ref{sec:borrowing}.}
    \item \mytextcolor{JSD: A borrowing method based on the Jensen-Shannon divergence (see Section~\ref{sec:pp}).}. 
    \item {EXNEX}: A hierarchical prior $\theta_i = \log \left( \frac{p_i}{1 - p_i} \right)$: 
    \begin{equation}
    \theta_i \sim w_{i1} \mathcal{N}(\mu_{\text{ex},1}, \sigma_{\text{ex},1}^2) + w_{i2} \mathcal{N}(\mu_{\text{ex},2}, \sigma_{\text{ex},2}^2) + w_{i0} \mathcal{N}(\mu_{\text{nex},i}, \sigma_{\text{nex},i}^2),
    \end{equation}
    where $w_{i1}, w_{i2}, w_{i0} = (0.25, 0.25, 0.5)$, $\mu_{\text{ex},1} \sim \mathcal{N}(-1.73, 6.84)$, $\mu_{\text{ex},2} \sim \mathcal{N}(-0.85, 3.76)$, $\sigma_{\text{ex},1}^2, \sigma_{\text{ex},2}^2 \sim \text{Halfnormal}(0,1)$, $\mu_{\text{nex},i} = -1.24$ and $\sigma_{\text{nex},i}^2 = 5.73$.
    \item {BHM}: A hierarchical prior on $\theta_i = \log \left( \frac{p_i}{1 - p_i} \right)$:
    \begin{equation}
    \theta_i | \mu, \sigma^2 \sim \mathcal{N}(\mu, \sigma^2), \quad \mu \sim \mathcal{N}(0, 100), \quad \sigma^2 \sim \text{Uniform}(0, 100).
    \end{equation}
    The uniform prior was used for $\sigma^2$ following the recommendation by \cite{cunanan2019variance}.
    \item {BCHM}: A hierarchical prior $\theta_i = \log \left( \frac{p_i}{1 - p_i} \right)$:
    \begin{equation}
    \theta_i | \mu, \sigma^2 \sim \mathcal{N}\left(\mu, \frac{1}{\tau^2 C_{ij}}\right), \quad \mu \sim \mathcal{N}(-1.73, 100), \quad \tau^2 \sim \text{Gamma}(50, 10),
    \end{equation}
    where $C_{ij}$ is the probability of baskets $i$ and $j$ being classified into the same cluster, estimated using Dirichlet process mixture \citep{neal2000markov}. The hyperparameters involved estimating $C_{ij}$ are set to $\sigma_0^2 = 10, \alpha = 10^{-40}, d_0 = 0, \sigma^2 = 0.001$; see \cite{chen2020bayesian} for these notation definitions.
    \item {local-MEM}: The original paper considered $\delta = 0, 1, 2$ and showed that the method with $\delta = 2$ keeps both family-wise and basket-wise type I error rates under control. Therefore, we set $\delta = 2$ for local-MEM.
    \item {MEM}: This method can be fit using the R package \texttt{basket} via the exact method. 
\end{itemize}
For all methods involving beta priors, a $\Beta(0.15, 0.85)$ prior is used for $p_i$, providing a prior mean equal to the null hypothesis and the prior information equivalent to one subject. There are no tuning parameters for PP-PEB, PP-GEB and IM. \mytextcolor{Tuning parameters for EXNEX, BHM, BCHM, local-MEM and MEM are set to the default values recommended by the original authors, as additional tuning is computationally intensive}. Since EXNEX has been observed to perform similarly to various newly proposed methods in the literature \citep[e.g.,][]{zhou2021robot,broglio2022comparison,lyu2023muce}, we focus on a detailed comparison of power prior based methods to EXNEX. 

\mytextcolor{In Bayesian basket trial designs, selecting appropriate tuning parameters is crucial for balancing power and type I error control.} \mytextcolor{For local-PP and JSD methods, we consider two tuning strategies:}

\mytextcolor{\textbf{Strategy 1: Optimized for General Evaluation}}. This approach selects tuning parameters that maximize power in terms of TPR and CCR averaged across Scenarios S2-S5, \mytextcolor{while ensuring that type I error inflation remains below 0.2 (i.e., the maximum BWER under Scenarios S2-S5 is controlled below 0.2)}. The threshold of 0.2 is used here as an illustrative example. For the proposed local-PP-PEB and local-PP-GEB methods, we select $a$ from $[0, 4]$ and $\Delta$ from $[0.1, 0.4]$. Here, 4 is the maximum borrowing factor, as $BF_i\leq n_{-i}/n_i =4$ with $n_i=25$ for all $i$. We use a maximum $\Delta$ of $0.4$ \mytextcolor{based on the assumption that} no borrowing should occur if the observed ORR difference exceeds 0.4. \mytextcolor{However, this upper bound can be adjusted depending on the trial context.} The selected tuning parameters are $(a=0.9,\Delta=0.4)$ for local-PP-PEB, and are $(a=3,\Delta=0.4)$ local-PP-GEB. The same tuning is applied to the JSD method \citep{fujikawa2020bayesian}, with $\epsilon$ selected from $[1,7]$ and $\tau$ from $\{0, 0.1, \ldots, 1\}$, and the resulted tuning parameters are $(\epsilon=3, \tau=0.5)$.

\mytextcolor{\textbf{Strategy 2: Tuned to Match EXNEX}}. \mytextcolor{Since EXNEX serves as a widely used benchmark model}, we tune the local-PP-PEB, local-PP-GEB and JSD methods to match EXNEX in terms of type I error inflation. \mytextcolor{This ensures that differences in power reflect the borrowing mechanisms rather than disparities in type I error control.} The resulting parameters are $(a=0.35, \Delta=0.4)$ for local-PP-PEB, $(a=0.45, \Delta=0.4)$ for local-PP-GEB, and are $(\epsilon=6.5, \tau=0.5)$ for JSD.

\mytextcolor{By presenting two sets of tuning parameters, we offer a comprehensive evaluation of the local-PP framework. The first set demonstrates its general performance, while the second set ensures a fair comparison with EXNEX by aligning type I error control. The results for both configurations are summarized in Table~\ref{tab:performance}, where the fine-tuned versions of local-PP-PEB, local-PP-GEB, and JSD are specifically labeled to indicate their alignment with EXNEX’s type I error control.}

We have provided a freely available R package, \texttt{BasketTrial}, for implementing the IM, JSD, PP-PEB, PP-GEB, local-PP-PEB, and local-PP-GEB methods, as well as for evaluating their operating characteristics via simulations. The package can be accessed at \url{https://github.com/wonderzhm/BasketTrial}. The R code to reproduce all results presented in this work is available at \url{https://github.com/wonderzhm/localPP}.  All R code was executed in R version 4.4.1 under the x86\_64-w64-mingw32/x64 (64-bit) platform. Computation time for each method was recorded using the actual running time recorded by the R function \texttt{Sys.time()}, utilizing 10 cores for parallel computing.

\subsection{Results}\label{sec:simu:results}
\textbf{Overall performance.}
The overall performance of all considered methods is summarized in Table~\ref{tab:performance}. Note that the same maximum sample sizes $n_i=25$ are assumed for all baskets, which implies that theoretically, the efficacy cutoff $Q_i$ should also be the same across baskets. Therefore, a common cutoff $Q_i=Q$ is calculated following the procedure outlined in Appendix A2. 

\begin{table}[h]
\centering
\caption{Overall performance of different methods \mytextcolor{under the equal basket size setting}. FPR is the average basket-wise type I error under the global null. BWER-avg is the average basket-wise type I error rate for non-promising baskets across scenarios S1-S5. BWER-max is the maximum basket-wise type I error rate for non-promising baskets across scenarios S1-S5. TPR-avg is the average of true positive rate across scenarios S2-S6. CCR-avg is the average of correct classification rate across scenarios S2-S6. Time is measured in hours. Tuning parameters are $(a=0.9, \Delta=0.4)$ for local-PP-PEB1, $(a=3,\Delta=0.4)$ for local-PP-GEB1, $(\epsilon=3, \tau=0.5)$ for JSD1, $(a=0.35,\Delta=0.4)$ for local-PP-PEB2, $(a=0.45,\Delta=0.4)$ for local-PP-GEB2, and $(\epsilon=6.5, \tau=0.5)$ for JSD2. }
\label{tab:performance}
\begin{tabular}{lccccccc}
\hline
Method & $Q$ & FPR & BWER-avg & BWER-max & TPR-avg & CCR-avg & Time \\
\hline
  IM & 0.857 & 0.064 & 0.063 & 0.067 & 0.724 & 0.779 & 0.002 \\ 
  PP-PEB & 0.919 & 0.099 & 0.184 & 0.308 & 0.846 & 0.830 & 0.002 \\ 
  local-PP-PEB1 & 0.888 & 0.096 & 0.132 & 0.197 & 0.819 & 0.830 & 0.002 \\ 
  PP-GEB & 0.928 & 0.100 & 0.145 & 0.208 & 0.828 & 0.831 & 0.002 \\ 
  local-PP-GEB1 & 0.926 & 0.099 & 0.139 & 0.198 & 0.825 & 0.830 & 0.002 \\ 
  JSD1 & 0.939 & 0.100 & 0.130 & 0.196 & 0.813 & 0.827 & 0.004 \\ 
  EXNEX & 0.865 & 0.100 & 0.118 & 0.143 & 0.804 & 0.823 & 2.083 \\ 
  BHM & 0.864 & 0.100 & 0.158 & 0.253 & 0.835 & 0.832 & 0.379 \\ 
  BCHM & 0.874 & 0.100 & 0.116 & 0.155 & 0.795 & 0.817 & 1.702 \\ 
  local-MEM & 0.867 & 0.100 & 0.107 & 0.123 & 0.782 & 0.811 & 0.072 \\ 
  MEM & 0.920 & 0.100 & 0.208 & 0.379 & 0.852 & 0.825 & 12.65 \\ 
\hline
\multicolumn{8}{c}{Tuned to match EXNEX performance}\\
  local-PP-PEB2 & 0.857 & 0.100 & 0.118 & 0.143 & 0.805 & 0.824 & 0.002 \\ 
  local-PP-GEB2 & 0.871 & 0.102 & 0.120 & 0.143 & 0.806 & 0.824 & 0.002 \\ 
  JSD2 & 0.919 & 0.100 & 0.110 & 0.141 & 0.790 & 0.816 & 0.004 \\ 
\hline
\end{tabular}
\end{table}

First, all methods except for IM maintain the FPR (i.e., average type I error) under global null close to the target level of 0.1, demonstrating successful calibration of $Q_i$. {In contrast, IM has a notably lower type I error than the target level due to the discrete nature of BWER values in the absence of information borrowing across baskets. For example, a small change in $Q$ from 0.857 to 0.856 results in the FPR shifting from 0.064 to 0.138. Consequently, the comparison between the IM model and other borrowing methods is influenced by this discrepancy in type I error control.} Second, all methods with information borrowing exhibit inflated BWER (i.e., BWER-max $>$ 0.1), with local-MEM showing the least inflation and MEM showing the highest, while IM shows no BWER inflation. Third, all methods with information borrowing demonstrate significantly higher TPR-avg (i.e., average power) compared to IM, with MEM achieving the highest power. Fourth, given the trade-off between BWER-avg and TPR-avg, CCR-avg can be viewed as a metric that balances both, representing overall performance. We observe that all borrowing methods achieve much higher CCR-avg than IM, with BHM demonstrating the best overall performance, followed by PP-PEB, local-PP-PEB1, PP-GEB, and local-PP-GEB1. Lastly, the PP methods with EB-based weights are the fastest among all borrowing methods, completing simulations in under 10 seconds. By comparison, EXNEX takes 2.2 hours, and MEM takes 12.65 hours to complete the full simulation. This difference is crucial for practical implementation when evaluating operation characteristics via simulations.

{Notably, the local-PP methods under the 3-component framework (local-PP-PEB1 and local-PP-GEB1) offer better control in type I error inflation than the PP methods using unadjusted weights (PP-PEB and PP-GEB), even though both approaches achieve the same overall performance in terms of CCR-avg. Before applying the 3-component framework, PP-PEB shows higher TPR-avg and higher BWER-max than PP-GEB, indicating that PP-PEB allows for a wider range of borrowing. However, after applying the 3-component framework, the two methods perform very similarly to each other.}

Thanks to its efficient computation, a key advantage of the local-PP methods is the ability to tune model parameter significantly faster than the MCMC-based methods. For instance, the local-PP-PEB2 and local-PP-GEB2 methods are able to be tuned to match the BWER-max of EXNEX at 0.143. For JSD2, the closest achievable BWER-max is 0.141. After aligning type I error inflation, local-PP-PEB2 and local-PP-GEB2 perform very similarly to EXNEX in terms of TPR-avg and CCR-avg, and they slightly outperform JSD2.

\textbf{Performance by Each Scenario.} 
Here, we focus on comparing the performance across scenarios for IM, local-PP-PEB $(a=0.35, \Delta=0.4)$ and local-PP-GEB $(a=0.45, \Delta=0.4)$, JSD $(\epsilon=6.5, \tau=0.5)$ and EXNEX as displayed in Table \ref{tab:results}. This includes the basket-wise rejection rates (i.e., basket-wise type I errors for non-promising baskets and basket-wise powers for promising baskets), as well as trial-wise FPR, FDR, TPR, and CCR. The results for other methods are summarized in Appendix Table A2.

\begin{table}[h]
\centering
\small
\caption{\mytextcolor{Performance by each scenario under the equal basket size setting.} Summary of Basket-wise rejection rates, false positive rates (FPR), false discovery rates (FDR), true positive rates (TPR), and correct classification rates (CCR). NA means not applicable. Tuning parameters are $(a=0.35, \Delta=0.4)$ for local-PP-PEB, $(a=0.45, \Delta=0.4)$ for local-PP-GEB, and are $(\epsilon=6.5, \tau=0.5)$ for JSD.}
\label{tab:results}
\begin{tabular}{cccccccccc}
\hline
Method & \multicolumn{5}{c}{Type I Error / Power} & FPR & FDR & TPR & CCR \\
\cline{2-6}
 & Basket 1 & Basket 2 & Basket 3 & Basket 4 & Basket 5 \\
\hline
\multicolumn{10}{c}{\textbf{Scenario S1 (0.15, 0.15, 0.15, 0.15, 0.15)}}\\
  IM & 0.065 & 0.066 & 0.062 & 0.059 & 0.067 & 0.064 & 0.283 & NA & NA \\ 
  local-PP-PEB & 0.098 & 0.107 & 0.098 & 0.094 & 0.104 & 0.100 & 0.347 & NA & NA \\ 
  local-PP-GEB & 0.101 & 0.110 & 0.099 & 0.096 & 0.106 & 0.102 & 0.347 & NA & NA \\ 
  JSD & 0.099 & 0.108 & 0.094 & 0.094 & 0.105 & 0.100 & 0.324 & NA & NA \\ 
  EXNEX & 0.099 & 0.106 & 0.098 & 0.093 & 0.102 & 0.100 & 0.354 & NA & NA \\ 
\hline
\multicolumn{10}{c}{\textbf{Scenario S2 (0.15, 0.15, 0.15, 0.30, 0.30)}}\\
  IM & 0.065 & 0.060 & 0.065 & 0.621 & 0.626 & 0.063 & 0.092 & 0.623 & 0.811 \\ 
  local-PP-PEB & 0.133 & 0.128 & 0.134 & 0.725 & 0.727 & 0.131 & 0.154 & 0.726 & 0.811 \\ 
  local-PP-GEB & 0.136 & 0.131 & 0.135 & 0.730 & 0.731 & 0.134 & 0.157 & 0.731 & 0.812 \\ 
  JSD & 0.122 & 0.121 & 0.125 & 0.700 & 0.704 & 0.123 & 0.136 & 0.702 & 0.807 \\ 
  EXNEX & 0.130 & 0.127 & 0.133 & 0.721 & 0.723 & 0.130 & 0.152 & 0.722 & 0.811 \\ 
\hline
\multicolumn{10}{c}{\textbf{Scenario S3 (0.15, 0.30, 0.30, 0.30, 0.30)}}\\
  IM & 0.065 & 0.619 & 0.625 & 0.623 & 0.623 & 0.065 & 0.021 & 0.623 & 0.685 \\ 
  local-PP-PEB & 0.143 & 0.740 & 0.735 & 0.737 & 0.739 & 0.143 & 0.039 & 0.738 & 0.762 \\ 
  local-PP-GEB & 0.143 & 0.741 & 0.736 & 0.737 & 0.740 & 0.143 & 0.039 & 0.738 & 0.762 \\ 
  JSD & 0.141 & 0.727 & 0.721 & 0.723 & 0.725 & 0.141 & 0.037 & 0.724 & 0.751 \\ 
  EXNEX & 0.143 & 0.740 & 0.735 & 0.737 & 0.737 & 0.143 & 0.039 & 0.737 & 0.761 \\ 
\hline
\multicolumn{10}{c}{\textbf{Scenario S4 (0.15, 0.30, 0.30, 0.45, 0.45)}}\\
  IM & 0.062 & 0.613 & 0.631 & 0.955 & 0.958 & 0.062 & 0.016 & 0.789 & 0.819 \\ 
  local-PP-PEB & 0.131 & 0.722 & 0.750 & 0.970 & 0.973 & 0.131 & 0.031 & 0.854 & 0.857 \\ 
  local-PP-GEB & 0.131 & 0.723 & 0.750 & 0.970 & 0.973 & 0.131 & 0.031 & 0.854 & 0.857 \\ 
  JSD & 0.116 & 0.681 & 0.702 & 0.967 & 0.969 & 0.116 & 0.027 & 0.830 & 0.841 \\ 
  EXNEX & 0.131 & 0.722 & 0.750 & 0.970 & 0.973 & 0.131 & 0.031 & 0.854 & 0.857 \\ 
\hline
\multicolumn{10}{c}{\textbf{Scenario S5 (0.15, 0.45, 0.45, 0.45, 0.45)}}\\
  IM & 0.062 & 0.960 & 0.955 & 0.959 & 0.959 & 0.062 & 0.013 & 0.958 & 0.954 \\ 
  local-PP-PEB & 0.133 & 0.973 & 0.971 & 0.971 & 0.976 & 0.133 & 0.027 & 0.973 & 0.951 \\ 
  local-PP-GEB & 0.130 & 0.973 & 0.971 & 0.971 & 0.976 & 0.130 & 0.027 & 0.973 & 0.952 \\ 
  JSD & 0.088 & 0.965 & 0.961 & 0.963 & 0.965 & 0.088 & 0.018 & 0.964 & 0.953 \\ 
  EXNEX & 0.133 & 0.973 & 0.971 & 0.971 & 0.976 & 0.133 & 0.027 & 0.973 & 0.951 \\ 
\hline
\multicolumn{10}{c}{\textbf{Scenario S6 (0.30, 0.30, 0.30, 0.30, 0.30)}}\\
  IM & 0.627 & 0.632 & 0.625 & 0.612 & 0.629 & NA & NA & 0.625 & 0.625 \\ 
  local-PP-PEB & 0.733 & 0.740 & 0.741 & 0.724 & 0.744 & NA & NA & 0.737 & 0.737 \\ 
  local-PP-GEB & 0.734 & 0.741 & 0.741 & 0.725 & 0.744 & NA & NA & 0.737 & 0.737 \\ 
  JSD & 0.728 & 0.734 & 0.735 & 0.718 & 0.735 & NA & NA & 0.730 & 0.730 \\ 
  EXNEX & 0.733 & 0.740 & 0.741 & 0.724 & 0.744 & NA & NA & 0.737 & 0.737 \\ 
\hline
\end{tabular}
\normalsize
\end{table}

In Scenario S1, which represents the global null, all methods control the type I error rate at the 0.1 level, with the IM method being particularly conservative. In Scenarios S2-S6, the local-PP-PEB and local-PP-GEB methods consistently perform similarly to EXNEX across all evaluation metrics. In contrast, JSD exhibits slightly different performance across all scenarios. In Scenario S4, JSD shows a marginally lower type I error inflation for basket 1, but this is at the cost of significantly reduced power for baskets 2 \& 3. In Scenario S5, JSD on longer exhibits type I error inflation, as the weight $w_{1j}$ between basket 1 and other baskets is close to zero when $\epsilon=6.5$ and $\tau=0.5$. However, this improvement comes at the expense of lower power for other baskets. Finally, in Scenario S6, where all baskets are promising and type I error inflation is not a concern, JSD demonstrates the lowest power compared to local-PP-PEB and local-PP-GEB and EXNEX.

{For all other methods presented in Appendix Table A2, PP-PEB, PP-GEB, BHM and MEM have much higher basket-wise power, but at the cost of higher type I error inflation ($>0.2$); while BCHM and local-MEM show much lower type I error inflation, but at the cost of lower power. In contrast, our fine-tuned local-PP-PEB $(a=0.9, \Delta=0.4)$ and local-PP-GEB $(a=3,\Delta=0.4)$ methods show a good balance between type I error inflation and power in Scenarios S2-S4.}

\subsection{Performance under Unequal Sample Sizes}
In practice, the tumor baskets often do not have equal sample sizes at the time of analysis due to variations in disease prevalence and operational constraints. \mytextcolor{To evaluate the robustness of the proposed methdos in such settings}, we conducted additional simulations \mytextcolor{with final} basket sample sizes \mytextcolor{set to} $(n_1, \ldots, n_5) = (26, 16, 8, 17, 22)$. \mytextcolor{To maintain consistency with the equal sample size setting in Section~\ref{sec:scenarios}, we incorporated one interim futility analysis conducted after the first 10 enrolled subjects in each basket, following the same stopping rule: stop basket $i$ if the number of responses is less than or equal to 1. Since basket 3 has a maximum sample size of only 8, it does not undergo an interim futility analysis.} The following methods were compared: IM, local-PP-PEB $(a=\mytextcolor{0.55}, \Delta=0.4)$ and local-PP-GEB $(a=\mytextcolor{0.55}, \Delta=0.4)$, JSD $(\epsilon=6.5, \tau=0.5)$ and EXNEX. \mytextcolor{To ensure a fair comparison within the unequal sample size setting, tuning parameters for local-PP-PEB, local-PP-GEB, and JSD were re-optimized using the Strategy 2 approach described in Section~\ref{sec:sim:spec}, aligning their type I error inflation with EXNEX}. 

\begin{table}[H]
\centering
\caption{Summary of calibrated $Q_i$ \mytextcolor{under the unequal basket size setting} to ensure BWER $\leq 0.1$ under Scenario S1. Tuning parameters are $(a=\mytextcolor{0.55}, \Delta=0.4)$ for local-PP-PEB, $(a=\mytextcolor{0.55}, \Delta=0.4)$ for local-PP-GEB, and are $(\epsilon=6.5, \tau=0.5)$ for JSD.}
\label{tab:qi:unequal}
\begin{tabular}{lccccc}
\hline
Method & $Q_1$ & $Q_2$ & $Q_3$ & $Q_4$ & $Q_5$ \\
\hline
  IM & 0.835 & 0.816 & 0.914 & 0.784 & 0.798 \\ 
  local-PP-PEB & 0.884 & 0.874 & 0.890 & 0.866 & 0.880 \\ 
  local-PP-GEB & 0.886 & 0.873 & 0.890 & 0.865 & 0.881 \\ 
  JSD & 0.914 & 0.934 & 0.915 & 0.916 & 0.913 \\ 
  EXNEX & 0.855 & 0.850 & 0.904 & 0.830 & 0.854 \\ 
\hline
\end{tabular}
\end{table}

\mytextcolor{Unlike the equal sample size setting, unequal sample sizes necessitate basket-specific efficacy cutoffs $Q_i$, as shown in Table~\ref{tab:qi:unequal}. The overall performance of all considered methods is summarized in Table~\ref{tab:performance:unequal}. Local-PP-PEB and EXNEX exhibited comparable performance in terms of CCR-avg (both at 0.762), while local-PP-GEB and JSD (0.752 and 0.757, respectively) performed slightly worse.} 

\begin{table}[H]
\centering
\caption{Overall performance for different methods \mytextcolor{under the unequal basket size setting}. FPR is the average basket-wise type I error under the global null. BWER-avg is the average basket-wise type I error rate for non-promising baskets across scenarios S1-S5. BWER-max is the maximum basket-wise type I error rate for non-promising baskets across scenarios S1-S5. TPR-avg is the average of true positive rate across scenarios S2-S6. CCR-avg is the average of correct classification rate across scenarios S2-S6. Tuning parameters are $(a=\mytextcolor{0.55}, \Delta=0.4)$ for local-PP-PEB, $(a=\mytextcolor{0.55}, \Delta=0.4)$ for local-PP-GEB, and are $(\epsilon=6.5, \tau=0.5)$ for JSD.}
\label{tab:performance:unequal}
\begin{tabular}{lccccc}
\hline
Method & FPR & BWER-avg & BWER-max & TPR-avg & CCR-avg\\
\hline
  IM & 0.090 & 0.084 & 0.104 & 0.676 & 0.735 \\ 
  local-PP-PEB & 0.099 & 0.120 & 0.154 & 0.727 & 0.762 \\ 
  local-PP-GEB & 0.100 & 0.115 & 0.155 & 0.712 & 0.752 \\ 
  JSD & 0.091 & 0.108 & 0.150 & 0.714 & 0.757 \\ 
  EXNEX & 0.100 & 0.122 & 0.155 & 0.726 & 0.762 \\ 
\hline
\end{tabular}
\end{table}

Additionally, Table~\ref{tab:results1:unequal} provides a detailed comparison of performance across individual scenarios. Similar to the results observed under equal sample sizes, \mytextcolor{local-PP-PEB and EXNEX} methods demonstrated nearly identical performance across all scenarios and evaluation metrics, with a few exceptions. For instance, in Scenario S6, local-PP-PEB achieved significantly higher power for basket 5 compared to EXNEX. \mytextcolor{This improvement is likely due to local-PP-PEB's dynamic borrowing mechanism, which adjusts information borrowing based on the current basket's sample size, enabling more effective information sharing when sample sizes vary.} Conversely, in Scenario S4, local-PP-PEB displayed slightly lower power for basket 2 compared to EXNEX, \mytextcolor{suggesting a more conservative borrowing strategy for smaller baskets.} \mytextcolor{We also observe that local-PP-PEB outperforms local-PP-GEB in most scenarios. This difference highlights the impact of the empirical Bayes estimation strategy for similarity weights: PEB allows for more adaptive borrowing based on pairwise basket characteristics, such as sample size and response rate, while GEB applies more uniform borrowing weights based on pooled baskets, which may not be optimal in heterogeneous settings.} In contrast, JSD exhibited greater variability in performance across scenarios, sometimes outperforming other methods and, in other cases, underperforming. 

\begin{table}[H]
\centering
\caption{\mytextcolor{Performance by each scenario under the unequal basket size setting.} Summary of Basket-wise rejection rates, false positive rates (FPR), false discovery rates (FDR), true positive rates (TPR), and correct classification rates (CCR). NA means not applicable. Tuning parameters are $(a=\mytextcolor{0.55}, \Delta=0.4)$ for local-PP-PEB, $(a=\mytextcolor{0.55}, \Delta=0.4)$ for local-PP-GEB, and are $(\epsilon=6.5, \tau=0.5)$ for JSD.}
\label{tab:results1:unequal}
\begin{tabular}{cccccccccc}
\hline
Method & \multicolumn{5}{c}{Type I Error / Power} & FPR & FDR & TPR & CCR \\
\cline{2-6}
 & Basket 1 & Basket 2 & Basket 3 & Basket 4 & Basket 5 \\
\hline
\multicolumn{10}{c}{\textbf{Scenario S1 (0.15, 0.15, 0.15, 0.15, 0.15)}}\\
  IM & 0.075 & 0.081 & 0.104 & 0.089 & 0.098 & 0.090 & 0.379 & NA & NA \\ 
  local-PP-PEB & 0.099 & 0.100 & 0.099 & 0.098 & 0.100 & 0.099 & 0.366 & NA & NA \\ 
  local-PP-GEB & 0.100 & 0.100 & 0.100 & 0.101 & 0.100 & 0.100 & 0.375 & NA & NA \\ 
  JSD & 0.101 & 0.094 & 0.069 & 0.092 & 0.096 & 0.091 & 0.274 & NA & NA \\ 
  EXNEX & 0.099 & 0.100 & 0.100 & 0.100 & 0.100 & 0.100 & 0.384 & NA & NA \\ 
\hline
\multicolumn{10}{c}{\textbf{Scenario S2 (0.15, 0.15, 0.15, 0.30, 0.30)}}\\
  IM & 0.075 & 0.073 & 0.104 & 0.599 & 0.658 & 0.084 & 0.120 & 0.628 & 0.801 \\ 
  local-PP-PEB & 0.134 & 0.145 & 0.105 & 0.643 & 0.683 & 0.128 & 0.156 & 0.663 & 0.788 \\ 
  local-PP-GEB & 0.139 & 0.155 & 0.104 & 0.633 & 0.673 & 0.133 & 0.167 & 0.653 & 0.782 \\ 
  JSD & 0.115 & 0.150 & 0.137 & 0.633 & 0.665 & 0.134 & 0.146 & 0.649 & 0.779 \\ 
  EXNEX & 0.136 & 0.150 & 0.103 & 0.641 & 0.662 & 0.129 & 0.160 & 0.652 & 0.783 \\ 
\hline
\multicolumn{10}{c}{\textbf{Scenario S3 (0.15, 0.30, 0.30, 0.30, 0.30)}}\\
  IM & 0.074 & 0.535 & 0.448 & 0.597 & 0.653 & 0.074 & 0.025 & 0.558 & 0.632 \\ 
  local-PP-PEB & 0.154 & 0.662 & 0.449 & 0.681 & 0.723 & 0.154 & 0.047 & 0.629 & 0.672 \\ 
  local-PP-GEB & 0.150 & 0.672 & 0.449 & 0.645 & 0.685 & 0.150 & 0.048 & 0.613 & 0.660 \\ 
  JSD & 0.127 & 0.653 & 0.513 & 0.670 & 0.695 & 0.127 & 0.035 & 0.633 & 0.681 \\ 
  EXNEX & 0.155 & 0.680 & 0.448 & 0.699 & 0.680 & 0.155 & 0.047 & 0.627 & 0.670 \\ 
\hline
\multicolumn{10}{c}{\textbf{Scenario S4 (0.15, 0.30, 0.30, 0.45, 0.45)}}\\
  IM & 0.074 & 0.535 & 0.459 & 0.932 & 0.958 & 0.074 & 0.020 & 0.721 & 0.762 \\ 
  local-PP-PEB & 0.143 & 0.658 & 0.460 & 0.949 & 0.968 & 0.143 & 0.036 & 0.759 & 0.778 \\ 
  local-PP-GEB & 0.115 & 0.640 & 0.460 & 0.936 & 0.962 & 0.115 & 0.030 & 0.750 & 0.777 \\ 
  JSD & 0.110 & 0.577 & 0.468 & 0.931 & 0.959 & 0.110 & 0.026 & 0.734 & 0.765 \\ 
  EXNEX & 0.147 & 0.689 & 0.459 & 0.956 & 0.963 & 0.147 & 0.037 & 0.767 & 0.784 \\ 
\hline
\multicolumn{10}{c}{\textbf{Scenario S5 (0.15, 0.45, 0.45, 0.45, 0.45)}}\\
  IM & 0.074 & 0.910 & 0.777 & 0.932 & 0.959 & 0.074 & 0.017 & 0.894 & 0.901 \\ 
  local-PP-PEB & 0.147 & 0.952 & 0.777 & 0.953 & 0.969 & 0.147 & 0.032 & 0.913 & 0.901 \\ 
  local-PP-GEB & 0.104 & 0.938 & 0.777 & 0.936 & 0.962 & 0.104 & 0.023 & 0.903 & 0.902 \\ 
  JSD & 0.096 & 0.914 & 0.758 & 0.925 & 0.951 & 0.096 & 0.021 & 0.887 & 0.890 \\ 
  EXNEX & 0.149 & 0.958 & 0.777 & 0.961 & 0.963 & 0.149 & 0.033 & 0.915 & 0.902 \\ 
\hline
\multicolumn{10}{c}{\textbf{Scenario S6 (0.30, 0.30, 0.30, 0.30, 0.30)}}\\
  IM & 0.659 & 0.551 & 0.454 & 0.584 & 0.654 & NA & NA & 0.581 & 0.581 \\ 
  local-PP-PEB & 0.751 & 0.704 & 0.455 & 0.703 & 0.750 & NA & NA & 0.673 & 0.673 \\ 
  local-PP-GEB & 0.747 & 0.692 & 0.455 & 0.628 & 0.682 & NA & NA & 0.641 & 0.641 \\ 
  JSD & 0.722 & 0.680 & 0.543 & 0.683 & 0.718 & NA & NA & 0.669 & 0.669 \\ 
  EXNEX & 0.752 & 0.711 & 0.454 & 0.717 & 0.721 & NA & NA & 0.671 & 0.671 \\ 
\hline
\end{tabular}
\end{table}

In conclusion, the proposed \mytextcolor{local-PP-PEB method} performs comparably to EXNEX under unequal sample sizes, offering a flexible and efficient solution for managing information borrowing in real-world scenarios. \mytextcolor{Additionally, local-PP-PEB consistently outperforms local-PP-GEB in most scenarios, reinforcing the limitations of GEB-based weights when basket sizes are unequal.} However, the comparison with JSD is less clear due to its distinct behavior in terms of type I error control across different scenarios.

\section{Example}\label{sec:example}
In this section, we apply the proposed local-PP method to a basket trial designed to assess the effect of vemurafenib for treating nonmelanomas carrying the BRAF V600 variant, which has been previously analyzed in \citep{chen2023bayesian} using different information borrowing methods. Table~\ref{tab:braf} provides the number of responders, sample size, and response rate for each basket.

\begin{table}[h]
\centering
\caption{Summary of BRAF V600 study}
\label{tab:braf}
\begin{tabular}{cccc}
\hline
Tumor Type & Sample size & Number of Responses & Response rate \\
\hline
NSCLC & 19 & 8 & 0.421 \\
CRC vemu & 10 & 0 & 0 \\
CRC vemu+cetu & 26 & 1 & 0.038 \\
Bile duct & 8 & 1 & 0.125 \\
ECD or LCH & 14 & 6 & 0.429 \\
ATC & 7 & 2 & 0.286 \\
\hline
\end{tabular}
\end{table}

Following \citep{chen2023bayesian}, at the design stage, we set $p_{0} = 0.15$ for all baskets $i=1, \ldots, 6$ with sample sizes $(n_1, \ldots, n_6) = (19, 10, 26, 8, 14, 7)$, and control the basket-wise type I error rate at $\alpha = 0.05$. We then simulate $M = 100,000$ trials under the global null, without interim analyses, to calibrate the efficacy cutoff value $Q_i$ for each basket. {For illustration purpose, we present results using only the local-PP-PEB method, assuming the tuning parameters at the design design stage were set to $a=1$ and $\Delta=0.4$. This implies that the maximum borrowing amount is equal to each basket's sample size, and no borrowing occurs if the observed ORR difference exceeds 0.4. The resulting $Q_i$ values, type I errors, and posterior probabilities $P(p_i > 0.15 | \text{Data})$ are reported in Table~\ref{tab:efficacy}. Compared to the IM method, the local-PP-PEB method produces much higher posterior probability for ATC, which can be atributed to the estimated similarity matrix shown in Appendix Table A3. The ATC basket borrows 9\% of information from NSCLC, bile duct and ECD or LCH. In terms of posterior probabilities relative to the corresponding efficacy cutoffs, both NSCLC and ECD or LCH pass their efficacy boundaries under both the IM and local-PP-PEB methods. Although neither method claims efficacy for ATC, the local-PP-PEB method yields a much higher posterior probability than the IM method.}

\begin{table}[h]
\small
\centering
\caption{Efficacy cutoffs, type I errors, and posterior probabilities for IM and local-PP methods on BRAF V600 trial data when $\alpha = 0.05$.}
\label{tab:efficacy}
\begin{tabular}{c|cc|cc|cc}
\hline
Basket & \multicolumn{2}{c|}{$Q_i$} & \multicolumn{2}{c|}{Type I error} & \multicolumn{2}{c}{Posterior probability of $p_i > 0.15$} \\
\cline{2-7}
 & IM & local-PP-PEB& IM & local-PP-PEB & IM & local-PP-PEB \\
\hline
  NSCLC & 0.955 & 0.933 & 0.016 & 0.050 & 0.997 & 0.999 \\ 
  CRC vemu & 0.849 & 0.925 & 0.049 & 0.050 & 0.014 & 0.014 \\ 
  CRC vemu+cetu & 0.928 & 0.942 & 0.033 & 0.050 & 0.020 & 0.033 \\ 
  Bile duct & 0.915 & 0.908 & 0.021 & 0.050 & 0.332 & 0.324 \\ 
  ECD or LCH & 0.875 & 0.928 & 0.046 & 0.050 & 0.991 & 0.996 \\ 
  ATC & 0.943 & 0.930 & 0.013 & 0.049 & 0.761 & 0.879 \\ 
\hline
\end{tabular}
\normalsize
\end{table}

\section{Discussion and Conclusions}\label{sec:discussion}
We proposed a novel 3-component local power prior (local-PP) framework for information borrowing in exploratory basket trials. This framework, consisting of global borrowing control ($a$), pairwise similarity assessments ($s_{ij}$), and a borrowing threshold ($\Delta$), provides several significant advantages in terms of interpretability, flexibility, and computational efficiency when compared to traditional MCMC-based methods such as BHM and EXNEX. The local-PP framework offers a practical and intuitive approach to managing the extent of borrowing across heterogeneous tumor baskets, even in the presence of unequal sample sizes.  

The global borrowing parameter $a$ plays a crucial role in controlling the amount of information borrowed across all tumor baskets. It is designed to reflect the level of confidence in an experimental drug's potential to produce a shared tumor response across multiple baskets with similar molecular characteristics. Statistically, $a$ governs the maximum allowable borrowing from other tumor baskets for each individual basket, as defined by the borrowing factor introduced in Section~\ref{sec:borrowing}. For example, setting $a=1$ implies that the maximum number of subjects borrowed from other baskets equals the current basket's sample size. In situations where sample sizes are limited, particularly in rare tumor types or when tumor heterogeneity is a major concern, a customized borrowing parameter ($a_i$) can be introduced to control the borrowing for specific baskets. This flexibility allows for more precise borrowing control in cases where the global setting may not be appropriate. The threshold parameter $\Delta$ further ensures that borrowing is restricted when there are significant differences in response rates between baskets, addressing concerns of over-borrowing. 

We \mytextcolor{examined} two methods for estimating the pairwise similarity component $s_{ij}$: pairwise empirical Bayesian (PEB) and global empirical Bayesian (GEB). Although the unadjusted GEB weights (where $w_{ij}=s_{ij}$) have been recommended in both the multiple historical study setting \citep{gravestock2019power} and the basket trial setting \citep{baumann2024basket}, we observed non-intuitive borrowing behaviors, as discussed in Section~\ref{sec:borrowing}. GEB also demonstrated a narrower range of borrowing compared to PEB. After incorporating our proposed 3-component framework, both methods exhibited similar operating characteristics \mytextcolor{when basket sizes were equal. However, under unequal basket sizes, PEB weights consistently outperformed GEB weights across most scenarios, indicating that PEB facilitates more effective borrowing in settings with sample size imbalance.} Given these findings, we recommend using the PEB weights over GEB within the 3-component framework for information borrowing in exploratory basket trials. Additionally, other similarity measures, such as the Jensen-Shannon divergence \citep{fujikawa2020bayesian} or calibrated power prior weights \citep{baumann2024basket}, can also be integrated into the 3-component framework.

Our simulation results demonstrate that the local-PP framework performs comparably to other existing methods in terms of power, type I error control, and correct classification rates (CCR). The introduction of the global borrowing parameter $a$ and the threshold $\Delta$ allows for flexible tuning of the borrowing mechanism, which can be adapted based on the expected heterogeneity between tumor types. In scenarios with greater tumor heterogeneity, the ability to customize $a$ or introduce basket-specific parameters $a_i$ enables precise control over the borrowing amount. The local-PP method consistently demonstrated strong performance across a range of scenarios, achieving high TPR and CCR while maintaining acceptable type I error inflation, particularly in comparison to more complex methods like EXNEX. 

\mytextcolor{To ensure a fair comparison with other borrowing methods, we focus on EXNEX, as it has been shown to perform similarly to various newly proposed approaches. Our strategy calibrates local-PP-PEB, local-PP-GEB, and JSD to match EXNEX’s type I error inflation. Alternatively, if computational cost were not a concern, one could optimize each method’s tuning parameters for specific scenarios and then compare all methods under their respective optimal configurations. However, there is no universally optimal configuration, as the best choice depends on the specific trial setting. The appropriate level of borrowing and type I error control should be carefully tailored to each study, considering both statistical and clinical inputs.}

\mytextcolor{Our study focuses on comparing Bayesian borrowing methods, with independent model (IM) serving as a reference rather than a direct comparator. While IM naturally exhibits lower type I error due to the discreteness of the binomial distribution, borrowing methods generally improve power when baskets share some similarity. An alternative approach could calibrate all methods to IM’s type I error (e.g., 0.064), but this would prioritize error control over power. Since basket trials typically aim for a balance between the two, we aligned Bayesian methods with the nominal 0.1 level. Future work could explore the implications of tuning all methods to IM’s error rate, particularly in settings with greater basket heterogeneity.}

While basket trials have the potential to improve trial efficiency, extensive simulations are necessary to optimize operating characteristics by appropriately setting tuning parameters. It is essential to effectively communicate the operating characteristics and trade-offs of each design option to the study team, helping guide the optimization of a basket trial design. The proposed borrowing framework has a particular advantage of model interpretation and making it easier to explain to cross functional team members.

The proposed method is in the context of exploratory basket trials, with the primary goal of identifying promising tumor types for further study. Regulatory approvals based on single-arm basket trials have historically been granted for exceptional drugs in terminal disease settings. However, \mytextcolor{there} is growing support for the use of randomized basket trials, as recommended by the French Health Technology Assessment Group \citep{lengline2021basket}. For guidance on basket trials in a confirmatory setting, we refer readers to the works of \citep{chen2016statistical}, \citep{li2017estimation}, and \citep{beckman2016adaptive}.


\noindent {\bf{Conflict of Interest}}

\noindent {\it{The authors have declared no conflict of interest. }}


\bibliographystyle{apalike}
\bibliography{basket}

\begin{thebibliography}{}

\bibitem[Baumann et~al., 2024]{baumann2024basket}
Baumann, L., Sauer, L.~D., and Kieser, M. (2024).
\newblock A basket trial design based on power priors.
\newblock {\em Statistics in Biopharmaceutical Research}, (just-accepted):1--18.

\bibitem[Beckman et~al., 2016]{beckman2016adaptive}
Beckman, R.~A., Antonijevic, Z., Kalamegham, R., and Chen, C. (2016).
\newblock Adaptive design for a confirmatory basket trial in multiple tumor types based on a putative predictive biomarker.
\newblock {\em Clinical Pharmacology and Therapeutics}, 100(6):617--625.

\bibitem[Berry et~al., 2013]{berry2013bayesian}
Berry, S.~M., Broglio, K.~R., Groshen, S., and Berry, D.~A. (2013).
\newblock {B}ayesian hierarchical modeling of patient subpopulations: Efficient designs of phase ii oncology clinical trials.
\newblock {\em Clinical Trials}, 10(5):720--734.

\bibitem[Broglio et~al., 2022]{broglio2022comparison}
Broglio, K.~R., Zhang, F., Yu, B., Marshall, J., Wang, F., Bennett, M., and Viele, K. (2022).
\newblock A comparison of different approaches to {B}ayesian hierarchical models in a basket trial to evaluate the benefits of increasing complexity.
\newblock {\em Statistics in Biopharmaceutical Research}, 14(3):324--333.

\bibitem[Chen and Hsiao, 2023]{chen2023bayesian}
Chen, C. and Hsiao, C. (2023).
\newblock {B}ayesian hierarchical models for adaptive basket trial designs.
\newblock {\em Pharmaceutical Statistics}, 22(3):531--546.

\bibitem[Chen et~al., 2016]{chen2016statistical}
Chen, C., Li, X.~N., Yuan, S., Antonijevic, Z., Kalamegham, R., and Beckman, R.~A. (2016).
\newblock Statistical design and considerations of a phase 3 basket trial for simultaneous investigation of multiple tumor types in one study.
\newblock {\em Statistics in Biopharmaceutical Research}, 8(3):248--257.

\bibitem[Chen and Lee, 2020]{chen2020bayesian}
Chen, N. and Lee, J.~J. (2020).
\newblock {B}ayesian cluster hierarchical model for subgroup borrowing in the design and analysis of basket trials with binary endpoints.
\newblock {\em Statistical Methods in Medical Research}, 29(9):2717--2732.

\bibitem[Cunanan et~al., 2019]{cunanan2019variance}
Cunanan, K.~M., Iasonos, A., Shen, R., and G{\"o}nen, M. (2019).
\newblock Variance prior specification for a basket trial design using {B}ayesian hierarchical modeling.
\newblock {\em Clinical Trials}, 16(2):142--153.

\bibitem[Fuglede and T{\o}psoe, 2004]{fuglede2004jensen}
Fuglede, B. and T{\o}psoe, F. (2004).
\newblock Jensen-shannon divergence and hilbert space embedding.
\newblock In {\em International Symposium on Information Theory, 2004. ISIT 2004. Proceedings.}, pages 31--. IEEE.

\bibitem[Fujikawa et~al., 2020]{fujikawa2020bayesian}
Fujikawa, K., Teramukai, S., Yokota, I., and Daimon, T. (2020).
\newblock A {B}ayesian basket trial design that borrows information across strata based on the similarity between the posterior distributions of the response probability.
\newblock {\em Biometrical Journal}, 62(2):330--338.

\bibitem[Gravestock and Held, 2019]{gravestock2019power}
Gravestock, I. and Held, L. (2019).
\newblock Power priors based on multiple historical studies for binary outcomes.
\newblock {\em Biometrical Journal}, 61(5):1201--1218.

\bibitem[Hobbs and Landin, 2018]{hobbs2018bayesian}
Hobbs, B.~P. and Landin, R. (2018).
\newblock {B}ayesian basket trial design with exchangeability monitoring.
\newblock {\em Statistics in Medicine}, 37(25):3557--3572.

\bibitem[Hyman et~al., 2015]{hyman2015vemurafenib}
Hyman, D.~M., Puzanov, I., Subbiah, V., Faris, J.~E., Chau, I., Blay, J.-Y., Wolf, J., Raje, N.~S., Diamond, E.~L., Hollebecque, A., et~al. (2015).
\newblock Vemurafenib in multiple nonmelanoma cancers with braf v600 mutations.
\newblock {\em New England Journal of Medicine}, 373(8):726--736.

\bibitem[Ibrahim et~al., 2015]{ibrahim2015power}
Ibrahim, J.~G., Chen, M.-H., Gwon, Y., and Chen, F. (2015).
\newblock The power prior: Theory and applications.
\newblock {\em Statistics in Medicine}, 34(28):3724--3749.

\bibitem[Jiang et~al., 2021]{jiang2021optimal}
Jiang, L., Nie, L., Yan, F., and Yuan, Y. (2021).
\newblock Optimal {B}ayesian hierarchical model to accelerate the development of tissue-agnostic drugs and basket trials.
\newblock {\em Contemporary Clinical Trials}, 107:106460.

\bibitem[Kane et~al., 2020]{kane2020analyzing}
Kane, M.~J., Chen, N., Kaizer, A.~M., Jiang, X., Xia, A., and Hobbs, B.~P. (2020).
\newblock Analyzing basket trials under multisource exchangeability assumptions.
\newblock {\em The R Journal}, 12(2):342.

\bibitem[Lenglin{\'e} et~al., 2021]{lengline2021basket}
Lenglin{\'e}, E., Peron, J., Vanier, A., Gueyffier, F., Kouzan, S., Dufour, P., Guillot, B., Blondon, H., Clanet, M., Cochat, P., et~al. (2021).
\newblock Basket clinical trial design for targeted therapies for cancer: A french national authority for health statement for health technology assessment.
\newblock {\em The Lancet Oncology}, 22(10):e430--e434.

\bibitem[Li et~al., 2017]{li2017estimation}
Li, W., Chen, C., Li, X., and Beckman, R.~A. (2017).
\newblock Estimation of treatment effect in two-stage confirmatory oncology trials of personalized medicines.
\newblock {\em Statistics in Medicine}, 36(12):1843--1861.

\bibitem[Liu et~al., 2022]{liu2022bayesian}
Liu, Y., Kane, M., Esserman, D., Blaha, O., Zelterman, D., and Wei, W. (2022).
\newblock {B}ayesian local exchangeability design for phase ii basket trials.
\newblock {\em Statistics in Medicine}, 41(22):4367--4384.

\bibitem[Lyu et~al., 2023]{lyu2023muce}
Lyu, J., Zhou, T., Yuan, S., Guo, W., and Ji, Y. (2023).
\newblock {MUCE}: {B}ayesian hierarchical modelling for the design and analysis of phase 1b multiple expansion cohort trials.
\newblock {\em Journal of the Royal Statistical Society Series C: Applied Statistics}, 72(3):649--669.

\bibitem[Neal, 2000]{neal2000markov}
Neal, R.~M. (2000).
\newblock Markov chain sampling methods for dirichlet process mixture models.
\newblock {\em Journal of Computational and Graphical Statistics}, 9(2):249--265.

\bibitem[Neuenschwander et~al., 2016]{neuenschwander2016robust}
Neuenschwander, B., Wandel, S., Roychoudhury, S., and Bailey, S. (2016).
\newblock Robust exchangeability designs for early phase clinical trials with multiple strata.
\newblock {\em Pharmaceutical Statistics}, 15(2):123--134.

\bibitem[Pohl et~al., 2021]{pohl2021categories}
Pohl, M., Krisam, J., and Kieser, M. (2021).
\newblock Categories, components, and techniques in a modular construction of basket trials for application and further research.
\newblock {\em Biometrical Journal}, 63(6):1159--1184.

\bibitem[Psioda et~al., 2021]{psioda2021bayesian}
Psioda, M.~A., Xu, J., Jiang, Q., Ke, C., Yang, Z., and Ibrahim, J.~G. (2021).
\newblock {B}ayesian adaptive basket trial design using model averaging.
\newblock {\em Biostatistics}, 22(1):19--34.

\bibitem[Simon et~al., 2016]{simon2016bayesian}
Simon, R., Geyer, S., Subramanian, J., and Roychowdhury, S. (2016).
\newblock The {B}ayesian basket design for genomic variant-driven phase ii trials.
\newblock {\em Seminars in Oncology}, 43(1):13--18.

\bibitem[Zabor et~al., 2022]{zabor2022bayesian}
Zabor, E.~C., Kane, M.~J., Roychoudhury, S., Nie, L., and Hobbs, B.~P. (2022).
\newblock {B}ayesian basket trial design with false-discovery rate control.
\newblock {\em Clinical Trials}, 19(3):297--306.

\bibitem[Zhou et~al., 2017]{zhou2017bop2}
Zhou, H., Lee, J.~J., and Yuan, Y. (2017).
\newblock {BOP2}: {B}ayesian optimal design for phase ii clinical trials with simple and complex endpoints.
\newblock {\em Statistics in Medicine}, 36(21):3302--3314.

\bibitem[Zhou and Ji, 2021]{zhou2021robot}
Zhou, T. and Ji, Y. (2021).
\newblock {RoBoT}: A robust {B}ayesian hypothesis testing method for basket trials.
\newblock {\em Biostatistics}, 22(4):897--912.

\end{thebibliography}


\end{document}



 \centerline{\large\bf A Bayesian Basket Trial Design Using Local Power Prior}

\vspace{.25cm}
\centerline{Haiming Zhou$^a$, Rex Shen$^b$, Sutan Wu$^a$ and Philip He$^a$}
\vspace{.4cm}
 \centerline{\it $^a$Daiichi Sankyo, Inc;  $^b$Stanford University}
\vspace{.55cm}
 \centerline{\bf Supplementary Material}
\vspace{.55cm}
\fontsize{9}{11.5pt plus.8pt minus .6pt}\selectfont
\noindent
\par

\setcounter{section}{0}
\setcounter{equation}{0}
\def\theequation{A\arabic{section}.\arabic{equation}}
\def\thesection{A\arabic{section}}
\def\thefigure{A\arabic{figure}}
\def\thetable{A\arabic{table}}

\fontsize{12}{14pt plus.8pt minus .6pt}\selectfont

\section{FDA Agnostic Approvals 2017--2022}\label{Append:fda}
\setcounter{equation}{0}
This section provides additional information for Section 1 in the main paper.
\begin{table}[H]
\centering
\caption{FDA Agnostic Approvals 2017 – 2022}
\label{tab:fda}
\begin{tabular}{l|l|p{6cm}|l}
\hline
FDA Approval & Drug & Setting & ORR (N) \\
\hline
May 2017 & Pembrolizumab & Unresectable/metastatic MSI-H or dMMR solid tumors that progressed after previous treatment with no satisfactory alternative treatment options & 40\% (59) \\
\hline
November 2018 & Larotrectinib & Unresectable/metastatic NTRK gene fusion‒positive solid tumors without a known acquired resistance mutation that progressed after previous treatment or with no satisfactory alternative treatment options & 75\% (55) \\
\hline
August 2019 & Entrectinib & Unresectable/metastatic NTRK gene fusion‒positive solid tumors without a known acquired resistance mutation that progressed after previous treatment or with no satisfactory alternative treatment options & 57\% (54) \\
\hline
June 2020 & Pembrolizumab & Unresectable/metastatic TMB-H ($\geq10$ mut/Mb) solid tumors that progressed after previous treatment with no satisfactory alternative treatment options & 29\% (102) \\
\hline
August 2021 & Dostarlimab & Recurrent/advanced dMMR solid tumors that progressed on or after previous treatment with no satisfactory alternative treatment options & 42\% (209) \\
\hline
June 2022 & Dabrafenib+trametinib & Unresectable/metastatic solid tumors with BRAF-V600E mutation that progressed after previous treatment with no satisfactory alternative treatment options & 41\% (131) \\
\hline
September 2022 & Selpercatinib & Locally advanced/metastatic solid tumors with a RET gene fusion that progressed on or after previous treatment or with no satisfactory alternative treatment options & 44\% (41) \\
\hline
\end{tabular}
\end{table}

\section{Calibration of $Q_i$}
This section provides additional information for Section 2.3 in the main paper. Below, we describe the calibration method based on BWER, which is also applicable to FPR, FWER or FDR when deemed appropriate in a particular study.
\begin{enumerate}
\setlength\itemsep{0em}
    \item Simulate a large number ($M$) of trials (e.g., $M=10,000$) under the global null hypothesis.
    \item For each trial $j$, calculate $q_{ij}$, the final posterior probability of $p_i > p_{0}$ for basket $i$, $i=1, \ldots, B$.
    \begin{itemize}
    \setlength\itemsep{0em}
        \item If basket $i$ has an early futility stop at the $k$-th interim, there is zero probability that basket $i$ can be claimed promising, that is, we set $q_{ij} = 0$ in this case.
        \item If basket $i$ has no early futility stop, $q_{ij} = P(p_i > p_{0} | D_{ij}, i \in A_j)$, where $A_j$ denotes the set of baskets at the final look and $D_{ij}$ is the accumulated data for basket $i \in A_j$; note that information borrowing across baskets in $A_j$ is applied in this calculation.
    \end{itemize}
    \item Calculate $Q_i$ as the $(1-\alpha)$-th quantile of $\{q_{ij}\}_{j=1, \ldots, M}$.
\end{enumerate}
Additionally, when multiple baskets have the same sample size $n_i$, the same $Q_i$ should be used for these baskets. In this case, $Q_i$ is calculated as the $(1-\alpha)$-th quantile of $\{q_{ij}\}_{j=1, \ldots, M, i \in S}$, where $S$ is the set of baskets sharing the same $n_i$. After determining $Q_i$, the basket-wise power is also calculated by simulations. Suppose $\tilde{M}$ trials are simulated under the alternative hypotheses. Then, the power for basket $i$ is estimated by the proportion of trials that have the posterior probability $P(p_i > p_{0} | \{D_i, i \in A_j\}) > Q_i$.

\section{Additional Results for the Simulation}\label{Append:simu}
This section provides additional information for Section 3 in the main paper.
\subsection{Additional results for equal basket sizes}
\begin{table}[H]
\scriptsize
\centering
\caption{\scriptsize Summary of Basket-wise rejection rates, false positive rates (FPR), false discovery rates (FDR), true positive rates (TPR), and correct classification rates (CCR). NA means not applicable. Tuning parameters are $(a=0.9, \Delta=0.4)$ for local-PP-PEB, $(a=3,\Delta=0.4)$ for local-PP-GEB, and $(\epsilon=3, \tau=0.5)$ for JSD.}
\label{tab:results1}
\begin{tabular}{cccccccccc}
\hline
Method & \multicolumn{5}{c}{Type I Error / Power} & FPR & FDR & TPR & CCR \\
\cline{2-6}
 & Basket 1 & Basket 2 & Basket 3 & Basket 4 & Basket 5 \\
\hline
\multicolumn{10}{c}{\textbf{Scenario S1 (0.15, 0.15, 0.15, 0.15, 0.15)}}\\
  PP-PEB & 0.100 & 0.101 & 0.097 & 0.095 & 0.101 & 0.099 & 0.232 & NA & NA \\ 
  local-PP-PEB & 0.095 & 0.104 & 0.093 & 0.089 & 0.098 & 0.096 & 0.302 & NA & NA \\ 
  PP-GEB & 0.103 & 0.105 & 0.096 & 0.093 & 0.103 & 0.100 & 0.263 & NA & NA \\ 
  local-PP-GEB & 0.102 & 0.105 & 0.096 & 0.092 & 0.102 & 0.099 & 0.262 & NA & NA \\ 
  JSD & 0.100 & 0.106 & 0.095 & 0.095 & 0.104 & 0.100 & 0.311 & NA & NA \\ 
  BHM & 0.098 & 0.106 & 0.100 & 0.093 & 0.103 & 0.100 & 0.336 & NA & NA \\ 
  BCHM & 0.099 & 0.109 & 0.096 & 0.094 & 0.105 & 0.100 & 0.333 & NA & NA \\ 
  local-MEM & 0.097 & 0.109 & 0.096 & 0.095 & 0.103 & 0.100 & 0.354 & NA & NA \\ 
  MEM & 0.100 & 0.106 & 0.096 & 0.097 & 0.102 & 0.100 & 0.214 & NA & NA \\ 
\hline
\multicolumn{10}{c}{\textbf{Scenario S2 (0.15, 0.15, 0.15, 0.30, 0.30)}}\\
  PP-PEB & 0.226 & 0.220 & 0.217 & 0.733 & 0.732 & 0.221 & 0.206 & 0.732 & 0.760 \\ 
  local-PP-PEB & 0.161 & 0.153 & 0.158 & 0.723 & 0.726 & 0.157 & 0.168 & 0.724 & 0.796 \\ 
  PP-GEB & 0.179 & 0.176 & 0.178 & 0.738 & 0.738 & 0.178 & 0.182 & 0.738 & 0.788 \\ 
  local-PP-GEB & 0.171 & 0.169 & 0.171 & 0.735 & 0.737 & 0.170 & 0.176 & 0.736 & 0.792 \\ 
  JSD & 0.154 & 0.152 & 0.158 & 0.719 & 0.720 & 0.155 & 0.159 & 0.720 & 0.795 \\ 
  BHM & 0.173 & 0.168 & 0.169 & 0.734 & 0.732 & 0.170 & 0.177 & 0.733 & 0.791 \\ 
  BCHM & 0.126 & 0.122 & 0.131 & 0.711 & 0.711 & 0.126 & 0.142 & 0.711 & 0.809 \\ 
  local-MEM & 0.113 & 0.109 & 0.115 & 0.698 & 0.702 & 0.112 & 0.132 & 0.700 & 0.813 \\ 
  MEM & 0.289 & 0.285 & 0.288 & 0.734 & 0.731 & 0.287 & 0.250 & 0.733 & 0.721 \\ 
\hline
\multicolumn{10}{c}{\textbf{Scenario S3 (0.15, 0.30, 0.30, 0.30, 0.30)}}\\
  PP-PEB & 0.308 & 0.804 & 0.805 & 0.807 & 0.813 & 0.308 & 0.072 & 0.807 & 0.784 \\ 
  local-PP-PEB & 0.197 & 0.766 & 0.762 & 0.763 & 0.764 & 0.197 & 0.050 & 0.764 & 0.772 \\ 
  PP-GEB & 0.208 & 0.776 & 0.773 & 0.775 & 0.778 & 0.208 & 0.052 & 0.775 & 0.779 \\ 
  local-PP-GEB & 0.198 & 0.771 & 0.769 & 0.768 & 0.774 & 0.198 & 0.050 & 0.771 & 0.777 \\ 
  JSD & 0.196 & 0.760 & 0.756 & 0.761 & 0.762 & 0.196 & 0.048 & 0.760 & 0.769 \\ 
  BHM & 0.253 & 0.784 & 0.780 & 0.783 & 0.786 & 0.253 & 0.060 & 0.783 & 0.776 \\ 
  BCHM & 0.155 & 0.731 & 0.725 & 0.727 & 0.727 & 0.155 & 0.040 & 0.727 & 0.751 \\ 
  local-MEM & 0.123 & 0.709 & 0.705 & 0.707 & 0.714 & 0.123 & 0.033 & 0.709 & 0.742 \\ 
  MEM & 0.379 & 0.823 & 0.822 & 0.823 & 0.828 & 0.379 & 0.088 & 0.824 & 0.784 \\ 
\hline
\multicolumn{10}{c}{\textbf{Scenario S4 (0.15, 0.30, 0.30, 0.45, 0.45)}}\\
  PP-PEB & 0.296 & 0.791 & 0.814 & 0.974 & 0.976 & 0.296 & 0.064 & 0.889 & 0.852 \\ 
  local-PP-PEB & 0.161 & 0.735 & 0.759 & 0.972 & 0.974 & 0.161 & 0.037 & 0.860 & 0.856 \\ 
  PP-GEB & 0.187 & 0.751 & 0.775 & 0.973 & 0.974 & 0.187 & 0.042 & 0.868 & 0.857 \\ 
  local-PP-GEB & 0.170 & 0.746 & 0.773 & 0.972 & 0.973 & 0.170 & 0.039 & 0.866 & 0.859 \\ 
  JSD & 0.156 & 0.711 & 0.730 & 0.970 & 0.972 & 0.156 & 0.035 & 0.846 & 0.845 \\ 
  BHM & 0.234 & 0.770 & 0.793 & 0.973 & 0.976 & 0.234 & 0.052 & 0.878 & 0.855 \\ 
  BCHM & 0.136 & 0.693 & 0.713 & 0.968 & 0.969 & 0.136 & 0.031 & 0.836 & 0.841 \\ 
  local-MEM & 0.117 & 0.682 & 0.706 & 0.966 & 0.968 & 0.117 & 0.027 & 0.831 & 0.841 \\ 
  MEM & 0.316 & 0.797 & 0.814 & 0.974 & 0.976 & 0.316 & 0.068 & 0.890 & 0.849 \\ 
\hline
\multicolumn{10}{c}{\textbf{Scenario S5 (0.15, 0.45, 0.45, 0.45, 0.45)}}\\
  PP-PEB & 0.259 & 0.977 & 0.976 & 0.974 & 0.977 & 0.259 & 0.053 & 0.976 & 0.929 \\ 
  local-PP-PEB & 0.141 & 0.972 & 0.971 & 0.971 & 0.976 & 0.141 & 0.029 & 0.973 & 0.950 \\ 
  PP-GEB & 0.167 & 0.974 & 0.971 & 0.971 & 0.976 & 0.167 & 0.034 & 0.973 & 0.945 \\ 
  local-PP-GEB & 0.156 & 0.974 & 0.971 & 0.971 & 0.976 & 0.156 & 0.032 & 0.973 & 0.947 \\ 
  JSD & 0.111 & 0.968 & 0.964 & 0.966 & 0.968 & 0.111 & 0.023 & 0.967 & 0.951 \\ 
  BHM & 0.241 & 0.976 & 0.975 & 0.974 & 0.976 & 0.241 & 0.049 & 0.975 & 0.932 \\ 
  BCHM & 0.104 & 0.966 & 0.963 & 0.965 & 0.965 & 0.104 & 0.021 & 0.965 & 0.951 \\ 
  local-MEM & 0.094 & 0.965 & 0.962 & 0.964 & 0.965 & 0.094 & 0.019 & 0.964 & 0.952 \\ 
  MEM & 0.234 & 0.975 & 0.974 & 0.973 & 0.976 & 0.234 & 0.048 & 0.974 & 0.933 \\ 
\hline
\multicolumn{10}{c}{\textbf{Scenario S6 (0.30, 0.30, 0.30, 0.30, 0.30)}}\\
  PP-PEB & 0.827 & 0.825 & 0.829 & 0.815 & 0.834 & NA & NA & 0.826 & 0.826 \\ 
  local-PP-PEB & 0.777 & 0.779 & 0.779 & 0.762 & 0.781 & NA & NA & 0.776 & 0.776 \\ 
  PP-GEB & 0.785 & 0.785 & 0.787 & 0.772 & 0.791 & NA & NA & 0.784 & 0.784 \\ 
  local-PP-GEB & 0.777 & 0.778 & 0.780 & 0.764 & 0.783 & NA & NA & 0.776 & 0.776 \\ 
  JSD & 0.774 & 0.777 & 0.777 & 0.763 & 0.778 & NA & NA & 0.774 & 0.774 \\ 
  BHM & 0.809 & 0.807 & 0.810 & 0.795 & 0.811 & NA & NA & 0.807 & 0.807 \\ 
  BCHM & 0.732 & 0.742 & 0.741 & 0.721 & 0.735 & NA & NA & 0.734 & 0.734 \\ 
  local-MEM & 0.705 & 0.713 & 0.713 & 0.693 & 0.712 & NA & NA & 0.707 & 0.707 \\ 
  MEM & 0.841 & 0.840 & 0.841 & 0.832 & 0.847 & NA & NA & 0.840 & 0.840 \\ 
\hline
\end{tabular}
\normalsize
\end{table}

\section{Additional Results for the Example}\label{Append:example}
This section provides additional information for Section 4 in the main paper.
\begin{table}[H]
\centering
\caption{The estimated similarity matrix for local PP method on BRAF V600 trial data.}
\label{tab:similarity}
\begin{tabular}{ccccccc}
\hline
 & NSCLC & CRC vemu & CRC vemu+cetu & Bile duct & ECD or LCH & ATC \\
\hline
  NSCLC & 1.00 & 0.00 & 0.00 & 0.09 & 0.29 & 0.29 \\ 
  CRC vemu & 0.00 & 1.00 & 0.03 & 0.00 & 0.00 & 0.00 \\ 
  CRC vemu+cetu & 0.01 & 0.15 & 1.00 & 0.45 & 0.02 & 0.07 \\ 
  Bile duct & 0.01 & 0.01 & 0.11 & 1.00 & 0.01 & 0.11 \\ 
  ECD or LCH & 0.20 & 0.00 & 0.00 & 0.07 & 1.00 & 0.20 \\ 
  ATC & 0.09 & 0.00 & 0.00 & 0.09 & 0.09 & 1.00 \\ 
\hline
\end{tabular}
\end{table}
